**Author**

Valeriya Chasova

**Supervisors**

Paul-Antoine Hervieux and Norbert Schappacher

**Title**

***Bessel-Hagen on the extension of Noether's theorems***
***and their application to classical electromagnetism***

**Abstract**

This work analyses the 1921 article by Erich Bessel-Hagen entitled *Über die Erhaltungssätze der Elektrodynamik* ("On the conservation laws of electrodynamics"). The article is based on Noether's theorems, which were formulated by Emmy Noether in 1918 and concern consequences of symmetries of actions, including conservation laws. Bessel-Hagen firstly extends Noether's theorems to symmetries up to a divergence and next applies them to the $n$-body problem and to classical electromagnetism, especially to its conformal symmetries. The work explains the context in which Bessel-Hagen's article was written, the way his explanations proceeded and the significance of his results. It is argued in particular that Bessel-Hagen's article as much as Noether's are best considered as elements of Felix Klein's Erlangen programme. A close reading of Bessel-Hagen's text is also provided, as is its comparison with relevant works, including recent works by philosophers of physics Katherine Brading, Harvey Brown and Peter Holland. Such an approach helps to see where Bessel-Hagen's treatment is confused and to clarify it, as well as to estimate the advancement in our understanding of the relevant topics made within the last one hundred years.

# Introduction

The present work provides an analysis of Erich Bessel-Hagen's article on conservation laws in electrodynamics [Bessel-Hagen 1921][1]. This article is a study of variational principles in physics, i.e. of physical results derivable from symmetries of actions. It concentrates on the derivation of conservation laws, i.e. of statements that some quantities do not change through time, taking as a basis Emmy Noether's theorems formulated a few years beforehand [Noether 1918]. Noether's first theorem links finite symmetries of actions (depending on parameters) with some divergence relationships, while her second theorem links infinite symmetries of actions (depending on functions) with some dependences. These theorems, and most easily the first, can be used to derive conservation laws.

Noether originally formulated both theorems on a very general level and for strict symmetries of action only. In his article Bessel-Hagen first extends Noether's theorems to some non-strict finite and infinite symmetries, namely to symmetries up to a divergence (§1). Afterwards he uses these results to derive conservation laws in two contexts. Firstly, more as a toy example, he derives the conservation laws following from the inhomogeneous Galilean symmetries in the context of the *n*-body problem (§2). Secondly, he discusses the derivation of conservation laws for classical electromagnetism, including the conservation of electric charge and current, but especially the conservation laws derivable from conformal symmetries (§§3-7). Though Bessel-Hagen relies on [Noether 1918] for §1, [Engel 1916] for §2 and a series of 1909-1910 works by Harry Bateman and Ebenezer Cunningham for §§3-7 ([Bateman 1909], [Bateman 1910a], [Bateman 1910b], [Cunningham 1910]; see [Kastrup 2008, pp. 632-634] on the relationships between these texts), as far as I can tell all of Bessel-Hagen's main results had not appeared in print anywhere before his article, and at least in part were obtained by himself.

Despite its rather rich content, Bessel-Hagen's article is not amongst the most known and analysed, and this by a number of reasons. In particular, Bessel-Hagen himself was a relatively marginal figure in his domain (to which his shyness and health problems contributed); the article (concerned with physics) was outside the author's main specialisation (mathematics and history of mathematics); the content of the article was derivative of more famous results (as it provides an extension and some applications of Noether's theorems); and its translation from German into English apparently came out late and in a not well known venue (at least the translation I am aware of, and was using together with the original for the present work, dates to 2006 and was published in some Swedish book series [Bessel-Hagen 1921/2006]).

---

[1]  Here and below, where references to a paragraph (§) without a year and author are given, the paragraphs are from Bessel-Hagen's 1921 article. His paragraphs are sections in the contemporary sense, taking a page or several pages each. The article contains an unnumbered introduction and 7 numbered paragraphs in total.

Yet by several reasons Bessel-Hagen's article does deserve attention. First, by extending the kind of symmetries Noether's theorems apply to it greatly extends the number of physical contexts where these theorems can in principle be used. Besides, it concentrates on the conformal symmetries, which is both rather original and relevant to certain approaches in physics, from Hermann Weyl's [1918] gauge theory to modern research in shape dynamics. More generally, and perhaps most importantly, it shows how to connect Noether's abstract mathematical treatment with physicists' practice. He therefore contributes in saving these theorems from the fate of many other potentially useful mathematical resources which due to the lack of similar demonstrations remain unapplied.

In the present work I will analyse Bessel-Hagen's article from both the historical and the conceptual perspective. Starting with the historical side, I will situate the article with respect to Bessel-Hagen's own life, the previous results he was relying on, and especially Felix Klein's Erlangen programme. There I will argue that Bessel-Hagen's article can be considered as one of the concluding steps of Klein's activity arising from that programme, and more radically that also Noether's theorems are just a step, notable as it is, in that programme led by Klein, despite the recent accent on Noether's role as expressed in particular by the influential book of Yvette Kosmann-Schwarzbach [2011]. Afterwards I will concentrate on the conceptual side of Bessel-Hagen's article, itself divided into a more mathematical part, concerning the extension of Noether's theorems, and a more physical part, concerning the application of the extended theorems to Bessel-Hagen's two case studies. Here I will give a detailed review of the article's content, its reformulation in contemporary terms where possible, and a comparison of it with some relevant works, including [Noether 1918], but also the recent works by philosophers of physics Harvey R. Brown, Katherine Brading and Peter Holland (such as [Brading and Brown 2000] and [Brown and Holland 2004]). While conceptual, such a discussion will feature a historical dimension as well, because it will allow to contrast the 100-year-old and current perspectives on the same questions, and thus to estimate how much our understanding of these questions has advanced in the meantime.

# Historical aspects

## *Bessel-Hagen's biography and the place of his article in it*

To start with, it is worth presenting Bessel-Hagen's biography and the place of his 1921 article there. Though Bessel-Hagen's life and work do not seem to be well-studied, here are some useful informations on them that can mostly be found, in German, in the beginning of [Neuenschwander 1993].

Erich Bessel-Hagen was born in 1898 in a well-educated family: one of his great grandfathers F. W. Bessel was a famous astronomer and the first director of Königsberg Observatory, while Erich's father F. C. Bessel-Hagen measured Immanuel Kant's skull on the occasion of Kant's burial in Königsberg, and was serving as a director of a hospital in Berlin. There also exists a so-called Bessel-Hagen decease, and though I could not find any reliable information on where the name comes from, one website suggests F. C. Bessel-Hagen was the first to describe it. My own speculation is that his son Erich deserves no less for giving his name to that decease if by chance he had it or to some other, as Erich's life was so affected by his bad health.

Yet he studied well, and by the end of a gymnasium in 1917 he excelled especially in Greek, physics and mathematics. Due to weak health and myopia, he was then able to stay in Berlin despite the continuing First World War. This allowed him to attend lectures by such specialists in his fields of interest as Ulrich von Wilamowitz-Moellendorff, Max Planck and Constantin Carathéodory (however I am not aware of his meeting Albert Einstein, though the latter was the director of Kaiser Wilhelm Institute for Physics in Berlin in 1917-1922). On 27 August 1920 Bessel-Hagen defended under the supervision of Carathéodory a thesis on variational problems in mathematics. In the autumn of the same year he moved to Göttingen, which at the time hosted plenty of strong mathematicians, including those having worked on the application of variational principles to physics: David Hilbert, Felix Klein and Emmy Noether. During the winter semester of 1920 a colloquium on the mathematical questions of general relativity took place, at which Klein suggested to apply Noether's theorems to classical electromagnetism. This suggestion is what led to Bessel-Hagen's 1921 article, as he reports in its introduction. The article was therefore written by Bessel-Hagen during his young years, at the time when his knowledge of physics was still fresh and his leaning towards the history of mathematics not yet predominant.

In 1921-1923 Bessel-Hagen continued as Klein's assistant, helping in particular to prepare the $3^{rd}$, final volume of Klein's collected works (however the $1^{st}$ volume that appeared in 1921 relates more to Bessel-Hagen's article, as it features the relevant texts [Klein 1910], [Klein 1918a], [Klein 1918b], [Klein 1918c], of which more below). Having obtained still in Göttingen a habilitation in mathematics in 1925, Bessel-Hagen took a teaching post in Halle in 1927 and moved in 1928 for another post in Bonn, where he eventually got a professorship and where he was to stay until the end of his life. During that time he apparently strengthened his previously developed interest in history of mathematics and even organised at the university a department devoted to that domain. Though reviewing and writing a lot, he was publishing little, either by his modest character or because of his still worsening health (Neuenschwander [1993, p. 385] reports him getting a hip fracture in 1929 and associated kidney problems especially from 1935 on).

At the time, unfortunately, both features were the reasons for mockery rather than sympathy, though apparently not as much in Bonn (as I am unaware of any stories from that period) as while Bessel-Hagen was younger and at less prestigious positions in Göttingen. In this respect the yet to be confirmed but way too often told story about almost just as young C. L. Siegel is most striking: he is reported to have thrown the only draft of Bessel-Hagen's habilitation thesis in the sea (Neuenschwander discusses various accounts of that story in [ibid., p. 388 and p. 409, n. 14]). Bessel-Hagen himself was by contrast a kinder person, helping in particular the Jewish mathematician Felix Hausdorff for some of the Nazi years, until Hausdorff and some of his family members were forced to commit a suicide in 1942. Bessel-Hagen's own death occurred in 1946, for while he was able to spend the Second World War in teaching due to his weak health, he could not survive the malnutrition and an outbreak of tuberculosis that followed. He left to future generations a considerable collection of lecture notes, letters and further materials by himself and others, which can now be found at the University of Bonn's archive; a description of these occupies the largest part of Neuenschwander's text [1993].

### Bessel-Hagen's and Noether's works
### as parts of Klein's Erlangen programme

Noether's theorems are decidedly better known than the use Bessel-Hagen would made of them, but they have long been known in a truncated form, namely one was most often referring to Noether's first theorem alone as a means of deriving conservation laws. There are many reasons for that truncated fame, of which at least three should be mentioned. Firstly, Noether made accent on deriving conservation laws from her first rather than second theorem, therefore those interested in conservation laws were not motivated to study the latter theorem. Secondly, though infinite (local) symmetries were already used in general relativity, they only became widespread after the blossoming of the gauge theories around half a century later, whence also a delay in appreciating

Noether's second theorem which deals with these symmetries. Thirdly, my speculation is that many simply did not know of Noether's second theorem and of the other contents of Noether's article as long as the English translation of it was remaining unavailable (thus apparently until 1971, see [Noether 1918/1971]) because of not knowing German, and that the proportion of such scholars was growing while the English-speaking scholarship was taking the lead over the German-speaking one.

In more recent times the situation started to change, however, so that by now mentioning Noether's two theorems is quite common, but there is something more. Namely, in reverence to the enduring usefulness of Noether's century-old results, and in compensation to Noether's difficulties as a woman during her life (particularly when trying to obtain a habilitation in Germany and a decent work status both there and in the USA) Noether has currently became a popular figure, and the appreciation of her theorems got elevated to a high extent. Kosmann-Schwarzbach's book [2011] on the reception of Noether's theorems across the XX century is characteristic of this change (and probably did much in bringing it in): its underlying thought is the utmost importance of Noether's work, and its content is the illustration of how the appreciation of that work was developing.

While doing justice to Noether is surely laudable, the current situation is unsatisfactory in that the works of her contemporaries now get overshadowed by her own. I say this not in view of claiming that Bessel-Hagen's work is to be appreciated more than Noether's, though for instance Bessel-Hagen's closer connection with physics is surely an advantage over Noether's abstract treatment when it comes to exhibiting the physical significance of her results. Instead, I wish to attract attention to a perspective from which both Noether's and Bessel-Hagen's results are just elements of a much more profound and comprehensive framework - this framework is the Erlangen programme of Felix Klein.

In 1872, at the age of just 23, Klein got a professorship at the University of Erlangen. Following a tradition, he made a programmatic inaugural speech there and soon published a text on its basis which became known as the Erlangen programme [Klein 1872]. The subsequent development of this programme is described by Klein in the $1^{st}$ volume of his collected works, in the beginning of the part containing his works that he takes to be related to the programme [Klein 1921, pp. 411-414]. Inspired by Klein's collaborations with Sophus Lie, the programme was initially centred on associating symmetry groups to spaces in geometry, but was not developing actively due to various practical reasons. However, a few decades later it got recast into the programme of applying symmetry groups to physics, particularly after Einstein formulated special relativity (1905) and Hermann Minkowski introduced a new space, now bearing his name, to give that theory a geometric interpretation (1907-1908). In the wake of this Klein wrote an article on the geometric foundations of the (inhomogeneous) Lorentz group, including its relationships with physics [Klein 1910]. His engagement with physics then became highly intensive due to the formulation of general relativity by Einstein [1915], the appearance of Hilbert's first note on the foundations of physics [1915] and the subsequent active discussion of the status of conservation laws in general relativity that involved Einstein, Hilbert, Klein and Noether (see e.g. [Einstein 1916], [Einstein 1918], [Rowe 1999], [Brading 2005]). These happenings led to several more publications by Klein, namely an extract of his correspondence with Hilbert [Klein 1918a] and two Klein's works respectively on the differential and the integral conservation laws [Klein 1918b], [Klein 1918c]. These last three works together with [Klein 1910] conclude the part of Klein's 1921 volume devoted to the Erlangen programme.

We thus see that for Klein his late Erlangen programme was an active period of research on the application of symmetry groups to physics, of which a study of conservation laws was an essential ingredient. Moreover, for Klein this research was by no means accidental, but on the contrary was rooted both in his past interest in the application of symmetry groups to geometry and in his overall attention to the application of mathematics to physics. That this attention was long-lasting is attested by Klein himself, which dates it back to the beginning of his mathematical studies ([1921, p. 413]), and by the huge *Encyclopedia of mathematical sciences including their applications*, whose publication Klein was co-directing in the last quarter of century of his life: the $5^{th}$ of its 6 volumes is devoted to physics, and notably contains Wolfgang Pauli's famous article on the relativity theory, published in 1921.

And on the other hand by contrast to this we have the works on symmetries and conservation laws in physics by Bessel-Hagen, Noether and Friedrich Engel (namely [Bessel-Hagen 1921], [Noether 1918], Noether's manuscript presumably from 1916 that Rowe [2019] tries to reconstruct, [Engel 1916] on which Bessel-Hagen's §2 relies very much, and [Engel 1917]). Besides the topic, there are at least two things that all these works have in common. Firstly, for all of the authors concerned these works were to my knowledge marginal in the sense that their authors were working their whole life predominantly on mathematics or history of mathematics without any further considerable interest in or contribution to the application of mathematics to physics. For Bessel-Hagen, in particular, his bibliography from the end of [Neuenschwander 1993] makes it clear that his 1921 article, with its accent on physics, is an exception from the range of his mathematical works. And secondly, all these works were in fact more or less directly inspired, and perhaps we can even say requested, by Klein himself. Indeed, for the works of Bessel-Hagen and Engel this is explicitly stated in their introductions. As to Noether's article, the decisive influence of Klein, as well as Hilbert, is perhaps a bit less evident from its text, but is very well attested and have been thoroughly analysed particularly in David Rowe's research (including [1999] and [2019]).

In light of this it should become obvious that the appearance of Bessel-Hagen's, Noether's and Engel's works was greatly due to the revival of Klein's Erlangen programme, rooted in Klein's research interests and stimulated by the relativistic advancements of Einstein, Minkowski and Hilbert. Therefore, while considering Bessel-Hagen's article as a follow-up on Noether's makes sense in a short-sighted perspective, which I will mostly concentrate on in what follows, one should also keep in mind a broader picture where both works are but parts of the vague of research brought about to a large extent by Klein.

### *On the originality of Bessel-Hagen's article*

While the young Bessel-Hagen got involved into Klein's movement and produced an article in the wake of Noether's results, that article was not completely unoriginal, as can be established by assessing its content, namely the extension of Noether's theorems to symmetries up to a divergence and its application to a couple of cases from physics.

The extension part is itself two-fold: one has to pass firstly from symmetries to symmetries up to a divergence, and secondly from Noether's theorems for the former to Noether's theorems for the latter symmetries. Bessel-Hagen describes both extensions in turn in his §1, and we find out that he credits Noether with both of them in the beginning of that section, because he says there that it is she who originally extended the theorems, and presumably one cannot extend the theorems without extending the notion of symmetry beforehand. One could then think that the material from §1 is completely unoriginal, but later in the same section, after having described the extension of the notion of symmetry and before passing to the extension of the theorems proper, Bessel-Hagen says that in the introduction of the new notion of symmetry up to a divergence is contained the extension that he was referring to in the beginning of the same section. So both his phrases converge on that Noether carried out the extension of the notion of symmetry, but the second phrase is less clear as to whether she extended the theorems as well. Moreover, it is not guaranteed that this can ever be established for sure, as Bessel-Hagen makes his attribution to Noether by referring to an oral communication of hers. Because of this it may remain unclear whether the extension part of Bessel-Hagen's article was in any way original with respect to Noether's reported contribution, though it is clear that at least to some extent it was not.

On the other hand, I am aware of no sources which would attribute to Noether, or indeed to anyone else prior to Bessel-Hagen, the application of Noether's theorems to the physics cases from Bessel-Hagen's §§2-7, so this application per se is anyway a novelty of Bessel-Hagen's article. However, Bessel-Hagen's description in §2 of conservation laws for the *n*-body problem happens to heavily rely on [Engel 1916], mentioned in his note 16. Thus of the application parts of Bessel-Hagen's article it is the remaining discussion of classical electromagnetism in §§3-7 which may be more original. Next, conformal symmetries of classical electromagnetism and some conservation laws of that theory were previously discussed in [Bateman 1910a] and in related works ([Bateman 1909], [Bateman 1910b] and [Cunningham 1910]) which Bessel-Hagen presumably refers to in his introduction, but as Kastrup remarks [2008, p. 633] none of these works connected conformal symmetries to new conservation laws. Nor did Weyl do so, we may add, when formulating his gauge theory in [1918], despite him going as far as deriving the conservation of electric charge from an infinite scale symmetry of action, which as Brading [2002] argues was an instance of applying Noether's second theorem. Thus Bessel-Hagen seems to be the first to explicitly combine Noether's theorems, conformal symmetries (understood as including a finite scale symmetry) and novel conservation laws (which are presented in his §7). This part of the content of his article seems therefore the most novel.

Returning to Noether, it is unclear why she had not written an article like Bessel-Hagen's herself. There were indeed enough reasons which could make her an ideal candidate for that: it was her who formulated the original theorems; presumably she also extended them, or the underlying notion of symmetry at the very least; she was well-versed at least in the general-relativistic aspects of physics, as can be seen from her discussion of it in [1918, §6] and in an earlier manuscript whose contents Rowe [2019] tries to reconstruct; and she should have been in Göttingen around the time when the colloquium took place and the article was written. Be that as it may, it is Bessel-Hagen who happened to write such an article, whether as an exercise or by other reasons. So it is he who should be praised for this work.

# Mathematical aspects

### *Noether's theorems and their physical understanding*

As Bessel-Hagen's article anyway relies essentially on Noether's [1918] original theorems, these theorems are worth being presented in quite some detail, as is the way their different elements are called, interpreted and used in physics.

Noether's theorems concern transformations. In general transformations may act on the *independent* or the *dependent* variables, where the latter variables are functions of the former. If a transformation acts on an independent variable, this *induces* a variation in a variable dependent on the variable acted upon. On the other hand, a dependent variable may also receive a *proper* variation if a transformation acts on that variable directly. Therefore in the case of dependent variables it makes sense to consider a *total* variation composed of an induced part and a proper part. By contrast, a transformation acting on a dependent variable cannot induce a change in any independent variable. Therefore, an independent variable has a proper variation alone. For the purpose of Noether's theorems both the transformations acting on the independent variables and the transformations acting on the dependent variables are considered.

On the other hand, for Noether's theorems any such transformations should be *symmetry transformations*, i.e. transformations leading to invariances. What has to be invariant under these transformations is more precisely an integral of a *functional*, where the latter means a function of a function. This functional may depend on the independent variables, the dependent variables and the derivatives of either kind of variables (so it is the dependence on the dependent variables and/or on their derivatives that makes it a functional). The integration of the functional is typically made over a domain specified in terms of the independent variables. For Noether's theorems the transformations may act in the bulk of the domain of integration and also at its boundaries.

In contemporary physics the relevant integral to be preserved is called an *action*, so I will call it in this way also when discussing Noether's and Bessel-Hagen's work, despite Noether usually speaking of just a certain integral and Bessel-Hagen of a variational problem instead (though Bessel-Hagen mentions the term 'action' at the very end of his article). In a typical particle context the functional is a Lagrangian, the independent variable is time and the dependent variables are spatial variables. Meanwhile, in a typical field context the functional is a Lagrangian density, the independent variables are spacetime variables and the dependent variables are field variables.

Here a *Lagrangian* and a *Lagrangian density* are quantities used to describe the evolution of a physical system (e.g. in mechanics a Lagrangian may encode the difference between kinetic and potential energy). The derivatives of these quantities feature in the so-called *Euler-Lagrange expressions*. *The Euler-Lagrange equations* are equations saying that the Euler-Lagrange expressions *vanish*, i.e. are equal to 0. From the physical point of view the Euler-Lagrange equations are equations of motion (for particles) or field equations (for fields) describing the evolution of physical systems.

One can derive the Euler-Lagrange equations from symmetries of actions via a variational principle called *Hamilton's principle* provided one considers transformations only of the dependent variables and only in the bulk of the domain of integration. This is by contrast with Noether's theorems, where one should also consider transformations of the independent variables and at the boundaries of the domain of integration. The *satisfaction* of the Euler-Lagrange equations, that is the fact for them to hold, is not needed to get the outcomes of Noether's theorems, but is needed to get from these outcomes *the usual conservation laws*. Whatever is valid prior to imposing the satisfaction of the Euler-Lagrange equations is said in mathematics to hold *identically*, so this can be said of the outcomes of Noether's theorems, for instance, but not of such conservation laws.

Noether considers not separate symmetry transformations, but their groups. A *group* in contemporary mathematics is a set equipped with a binary operation such that the set is closed under this operation, the operation is associative, there exist the identity element in the set, and there exists an inverse for every element in the set under the operation concerned. Noether as to her mentions the closure and the inverse in her informal definition from [ibid., §1]. Her set is a set of symmetry transformations, so its element is such a transformation, while its operation is a composition of such transformations, and the whole is a *symmetry group*.

She further divides these groups into *finite* continuous and *infinite* continuous according to whether the most general transformations in which these groups' transformations are contained depend respectively on a finite number of essential parameters, or on certain essential functions and their derivatives. In current physics and philosophy of physics literature finite and infinite groups are often named respectively *global* and *local*, although the global/local distinction also has other meanings there. Noether's first theorem concerns finite (global) groups, Noether's second theorem concerns infinite (local) groups.

*Noether's first theorem* says that from the invariance of an action under a finite symmetry group follow certain *divergence relationships*, and vice versa. These divergence relationships contain the Euler-Lagrange expressions on the left-hand side and a *divergence* of some quantity denoted *B* on the right-hand side, where the divergence of a quantity is a sum of derivatives of that quantity over the independent variables and over the terms depending on them. If one takes these divergence relationships and next imposes the satisfaction of the Euler-Lagrange equations, then the left-hand side vanishes and the result comes out as saying that the divergence of *B* vanishes. In particle theories one's search for conservation laws stops at this result; in the past such conservation laws were often called "first integrals", the name we encounter in Bessel-Hagen's article as well. In field theories, however, one may further integrate this result and impose suitable boundary conditions, in which case the former right-hand side transforms into the total time derivative of another quantity, and the result now comes out as saying that this last quantity is conserved. In field theories *B* and the other quantity are usually denoted respectively by *J* and *Q* (particularly in the electromagnetic case) and are usually called respectively a conserved current and a conserved charge. The expressions saying that *J* and *Q* are conserved are usually called *a continuity equation* and *a conservation law*, or alternatively a differential conservation law and an integral conservation law, respectively.

  *Noether's second theorem* says that from the invariance of an action under an infinite symmetry group follow certain *dependences*, and vice versa[2]. According to these dependences, a combination of the Euler-Lagrange expressions and of their derivatives vanishes. Such equations can therefore be interpreted as telling how these expressions and their derivatives depend on each other: this is apparently how Noether interprets them, as is seen from the way she names these equations. Alternatively, the same equations can be interpreted as *unusual conservation laws* provided they make a derivative of some quantity vanish: they are then unusual in so far as they hold identically[3]. Such conservation laws are known as *generalised Bianchi's identities*, because they amount to the so-called *Bianchi's identities* in the case of general relativity. The latter identities say that the Einstein tensor $G_{\mu\nu}$ vanishes without the need for any Euler-Lagrange equations to hold.

  Though Noether makes accent on deriving some usual conservation laws from her first theorem, there are actually many ways of deriving exactly the same conservation laws from her second theorem, as well as from other variational principles involving infinite symmetry groups, like the results of Utiyama [1959, p. 24]. These ways are described in [Brading and Brown 2000] and [Rowe 2002], in particular. Some of these ways require presupposing that an infinite symmetry group has a global subgroup (the case Noether discusses in [1918, §6]), but others do not. For instance, one of Brading and Brown's [2000] derivations of the conservation of electric current and charge in classical electromagnetism with matter only uses the generalised Bianchi's identities following from Noether's second theorem, plus the requirement that the gauge but not matter Euler-Lagrange expressions vanish. Meanwhile, in [Brading 2002, Appendix] the same conservation laws for *free* (i.e. without matter) classical electromagnetism are derived using the generalised Bianchi's identities alone, but following from Noether's second theorem as extended by Bessel-Hagen (though a direct reference to his work is not made).

---

2 I am calling these 'dependences' following M. A. Tavel's translation in [Noether 1918/1971], which is in agreement with the semantics of the original term 'Abhängigkeiten'. Meanwhile, Kosmann-Schwarzbach's translation of these as 'identities' in [Noether 1918/2011] is contrary to Noether's original, as Noether calls 'identities' ('Identitäten', in Kosmann-Schwarzbach's translation "identity relations") the outcomes of either theorem, see [Noether 1918, §1, p. 239]. The reason why Noether employs this qualification is apparently that either theorem holds identically in the sense above, i.e. prior to requiring the satisfaction of the Euler-Lagrange equations. Kosmann-Schwarzbach's translation choice is even less understandable given the appreciation of Noether's work that Kosmann-Schwarzbach tries to convey by her [2011].

3 An extensionally different way of defining the usual/unusual conservation laws distinction would be to divide them into those which can and those which cannot be obtained using Noether's first theorem. A yet another definition is apparently used by Bessel-Hagen: see on this my comments when discussing the *n*-body problem below.

Another confusion consists in thinking that conservation laws are determined by symmetries of the Euler-Lagrange equations or of Lagrangians and Lagrangian densities rather than by symmetries of actions. This confusion can be found in particular in the introduction of Bessel-Hagen's article, where it is said that Klein suggested to apply Noether's theorems to conformal symmetries of Maxwell's equations and that conservation laws should follow from that (whether this is Bessel-Hagen's confusion or also Klein's is less clear, because a similar formulation linking Noether's divergence relationships with conformal symmetries of Maxwell's equations is also found in Klein's 1921 letter to Pauli concerning Bessel-Hagen's article, translated in [Kosmann-Schwarzbach 2011, pp. 159-160]; Maxwell's equations are anyway an instance of the Euler-Lagrange equations). From the perspective of Noether's theorems and of the derivations of conservation laws proceeding via these theorems such thoughts are however erroneous, because the theorems only rely on symmetries of actions, and these symmetries need not be in one-to-one correspondence with symmetries of the Euler-Lagrange equations and of Lagrangians or Lagrangian densities. Indeed already the Euler-Lagrange equations, Lagrangians and Lagrangian densities, and actions themselves need not be in one-to-one correspondence with each other. Thus typically to the same Euler-Lagrange equations correspond multiple Lagrangians (see e.g. [Brown and Holland 2004a, Appendix A]) and multiple actions (see e.g. [ibid., *ii*)] and a clearer discussion in [Brown and Brading 2002, Sect. III]). Similarly, to given symmetries of some Euler-Lagrange equations correspond their presence and absence in the various actions which lead to these Euler-Lagrange equations (see e.g. the last two references and [Brown and Holland 2004b]) and in the various Lagrangians and Lagrangian densities which figure in these Euler-Lagrange equations (see e.g. [Brown and Holland 2004a, Appendix A, *iii*)]). And on the other hand, a transformation may change one Lagrangians or Lagrangian density into another while preserving the action, according to Kosmann-Schwarzbach's remark on symmetries up to a divergence discussed below.

### *Bessel-Hagen's extension of symmetries to symmetries up to a divergence*

In the very beginning of §1 Bessel-Hagen says that he is going to extend Noether's theorems from her article in the way Noether orally communicated to him. He then moves on to extending the notion of symmetry as follows.

First, he introduces the action $I$, the integrand $f$, the independent variables $x$ and the dependent variables $u$. He then says that $I$ is defined to be invariant under a one-to-one transformation of both kinds of variables provided $f$ is invariant under it.

Next, he introduces groups of infinitesimal transformations depending on the parameters $\varepsilon$ (for finite groups) or on the functions $p$ (for infinite groups). He then says that $I$ is defined to be invariant under these infinitesimal transformations provided $f$ is invariant under them at least in the terms which are first-order in $\varepsilon$, or respectively in $p$ and their derivatives.

Finally, he defines a divergence (denoted 'Div') of a quantity as a sum of its derivatives with respect to the independent variables and with respect to any terms dependent on them. He then says that $I$ is defined to be invariant up to a divergence under the previously considered infinitesimal transformations provided $f$ transforms under them into $f + \mathrm{Div}\,C$ + higher terms, where $C$ is an expression linear in $\varepsilon$, or respectively in $p$ and their derivatives, and may vanish identically in the trivial case. This is concluded by the note 11, which explains why the newly introduced notion requires concentrating on the infinitesimal transformations rather than on the corresponding one-parameter groups generated by these transformations.

Just after this extension of the notion of symmetry, and before passing to the extension of Noether's theorems themselves, is where Bessel-Hagen says that in the introduction of the new notion of symmetry up to a divergence is contained the extension, with respect to Noether's article, that he was referring to in the beginning of the same section. This phrase could be interpreted as ensuring a transition from the extension of the notion of symmetry to the extension of Noether's theorems, but the fact that the phrase continues a paragraph on symmetries up to a divergence does not square very well with that. Otherwise the same phrase can be interpreted as saying that Noether's orally communicated extension mentioned in the beginning of the section concerned the notion of symmetry alone and not her theorems as well.

### Comments on Bessel-Hagen's notion of symmetry up to a divergence
### and its relationship to Noether's work

Kosmann-Schwarzbach has an interesting note on how Bessel-Hagen's notion compares with Noether's: she says that Noether's condition of $f$ being unchanged amounts to the invariance of $f$, while Bessel-Hagen's condition of $f$ being changed into $f + \mathrm{Div}\, C$ amounts to the invariance of $I$ [2011, p. 92, n. 5]. As moreover the unchanged $f$ implies the unchanged $I$, we can infer from Kosmann-Schwarzbach's note that both Noether and Bessel-Hagen concentrate on the strict invariance of $I$, but Noether ensures this by requiring the strict invariance of $f$, while Bessel-Hagen allows this to also be ensured by the change of $f$ into $f + \mathrm{Div}\, C$. If this all is right, Bessel-Hagen's symmetries are actually symmetries of $f$ up to a divergence and not symmetries of $I$ up to a divergence, contrary to how he names them.

The real symmetries of $I$ up to a divergence, i.e. the cases where $I$ itself changes into $I + \int \mathrm{Div}\, D$ for some $D$, are mentioned in particular in the penultimate paragraph of [Noether 1918, §3]. This shows that extensions of symmetry-related notions were not unnatural for Noether, and other extension examples from her text go in the same direction (the case of the relative invariance of the left-hand side of her first theorem in [ibid., §1, n. 10]; the (apparently borrowed from physicists) passage from the vanishing Euler-Lagrange expressions to also the non-vanishing ones in [ibid., §1, n. 7 and §3, n. 15]). But none of these extensions is apparently the Noether's extension Bessel-Hagen was referring to in his §1, both because these extensions do not explicitly involve $f$ and because they were made in writing.

### Bessel-Hagen's extension of Noether's theorems to symmetries up to a divergence
### and its correspondence with Noether's article

Having introduced an extended notion of symmetry, Bessel-Hagen proceeds still in §1 to presenting the extension of Noether's theorems to the resulting kind of symmetries. His formulations of the extended theorems are simple: they say that the relevant divergence relationships and dependences follow for respectively finite and infinite symmetries up to a divergence, and that the converses hold as well.

When introducing these formulations, Bessel-Hagen also gives several equations, which we can divide in two categories according to whether they are valid for both theorems or for just one of them. He also refers to some sections and pages of Noether's article [1918] for an analogous treatment, but does not give more precise references to her equations. So I will now tell both of Bessel-Hagen's two kinds of equations and of their correspondences to Noether's equations that I have identified.

On the one hand, Bessel-Hagen gives four equations which follow after he had stated the extension of Noether's first theorem, but which should actually be valid for the extension of Noether's second theorem as well. This can be seen in particular from the absence of $\varepsilon$ and $p$ in these equations, as well as from the fact that Bessel-Hagen also refers to one of these equations, (5), after having passed to discussing Noether's second theorem. Noether as to her introduces the corresponding equations clearly before passing to either theorem.

Firstly, there is Bessel-Hagen's equation (5), which defines the total *infinitesimal variation $\delta$* of the dependent variables $u$ under the supposition that the domain of integration is unchanged. This corresponds to Noether's equation (9), with Bessel-Hagen's $\delta$ denoted by $\delta$-bar there, as opposed to Noether's own $\delta$ which does not require the supposition just mentioned. The introduction of the supposition is explained by Noether just above her equation (9).

Secondly, there is Bessel-Hagen's unnumbered equation just after his (5), which for Bessel-Hagen serves to define the quantity $A$, but which is actually the equation from which the divergence relationships are going to follow provided the symmetry group is finite. This corresponds to Noether's equation on the left of her (12), though this is worth also being compared with Noether's equation (3), defined however for Noether's own $\delta$ which is not Bessel-Hagen's.

Thirdly, there is Bessel-Hagen's equation (6), which expresses (in terms of components) the quantity $B$ as a sum involving $A$ and $\delta f$. This corresponds to Noether's equation on the right of her (12).

Fourthly, there is Bessel-Hagen's equation (8), which is what his (6) amounts to when $f$ depends on the derivatives of $u$ which are of first order only. The analogue of this equation seems missing from Noether's text, though she considers a similar case leading to her (4).

On the other hand, Bessel-Hagen gives five equations which are specific to one or the other of the theorems he considers. This can be seen from the fact that most of these equations feature $\varepsilon$ or $p$, which indicates that they presuppose a finite or an infinite symmetry group respectively, while the remaining equations presuppose that these equations featuring $\varepsilon$ or $p$ hold.

Three of these five equations concern Noether's first theorem. These are Bessel-Hagen's equation (7) and the unnumbered couple of equations preceding it. They correspond in Noether's text respectively to the equation (13) and to the unnumbered couple of equations preceding it (in the inverted order with respect to Bessel-Hagen's couple). In either case the unnumbered equations separate $\delta u$ and $B$ with respect to $\varepsilon$, so a finite symmetry group depending on $\varepsilon$ is now clearly presupposed. Meanwhile, the numbered equations say what Bessel-Hagen's previous unnumbered equation, corresponding to Noether's equation on the left of her (12), amounts to under the presupposition and the separation just mentioned. Namely, they amount to the divergence relationships arising from Noether's first theorem.

The two remaining Bessel-Hagen's equations, as to them, concern Noether's second theorem. One of these equations, (9), corresponds to Noether's equation just after her equation (13), provided we omit from Noether's corresponding equation the Euler-Lagrange expressions which both she and Bessel-Hagen denote by $\psi$. Finally, the other Bessel-Hagen's equation, (10), corresponds to Noether's equation (16). Both of the latter equations are the dependences arising from Noether's second theorem.

### *Comments on Bessel-Hagen's presentation of the relationship*
### *between (the extended) Noether's theorems and conservation laws*

It should be noted that several things are missing from Bessel-Hagen's presentation of the extended Noether's theorems. In particular, it is far less detailed than Noether's account in what concerns the derivations of the theorems, the conditions under which they hold and the like, though this is understandable in so far as the theorems are Noether's main topic and not Bessel-Hagen's. Much more surprising is Bessel-Hagen's complete silence over how in general Noether's theorems lead to conservation laws, which are however the main topic of Bessel-Hagen himself. This general account is missing from the general §1 where it would be the most appropriate, but moreover is not found elsewhere in the article either, except for a brief statement in the introduction, which is however very confusing.

Indeed, Bessel-Hagen says in the introduction that from symmetries of actions follows a number of relations which are satisfied identically by virtue of the differential equations. This is confusing if the equations are the Euler-Lagrange equations while "identically" means, as in my presentation above, "without the need for the Euler-Lagrange equations to be satisfied". He further says that these relations take the form of "first integrals" or conservation laws according to whether one independent variable or a finite group is involved. This is confusing in that the relations usually are not "first integrals" and conservation laws properly speaking but only turn to them after we require that the Euler-Lagrange equations be satisfied, and in addition to get conservation laws the integration has to be performed and suitable boundary conditions should hold. What Bessel-Hagen should have said is rather that from symmetries of actions a number of relations follow which are satisfied identically, but which can be transformed in "first integrals" or conservation laws typically with the help of the Euler-Lagrange equations.

Meanwhile, what we find in other sections of Bessel-Hagen's article is no less confusing, besides being restricted to specific cases. Thus in §2 and §6 Bessel-Hagen does impose the vanishing of the Euler-Lagrange expressions so as to get some continuity equations and conservation laws from the outcomes of Noether's theorems. However, he also discusses in §6 a case where in view of getting conservation laws the Euler-Lagrange expressions $\psi$ are assumed not to vanish but to be equal to some $P$ instead (compare this with one of Noether's extensions mentioned above). And above all, he deals in §§4, 5, 7 with finite and infinite symmetries alike. His discussion may therefore seem mysterious for someone who keeps from Noether's article the impression that conservation laws can primarily be derived from finite symmetries by imposing that the Euler-Lagrange expressions vanish.

Another thing missing from Bessel-Hagen's text is a discussion of what changes in the extended Noether's theorems as compared to the original ones, and how this impacts conservation laws derived using these theorems. We do have a few hints on this in his text, though. To begin with, it is clear that upon his extension the variation of $I$ may lead to the appearance of Div$C$. We note then that this $C$ carries over to Bessel-Hagen's equation (6): indeed the presence of $C$ is the crucial difference between this equation and Noether's corresponding equation on the right of her (12). Both equations define $B$, whose divergence figures on the right-hand side of the divergence relationships arising from Noether's first theorems. In other words, as already mentioned $B$ is a current that gets conserved once the Euler-Lagrange expressions on the left-hand side of the divergence relationships are assumed to vanish. Therefore, the impact of extending symmetries to symmetries up to a divergence is that the term whose divergence these symmetries make to appear contributes to the value of the current to be conserved.

Or at least this should be so for continuity equations and conservation laws derived using Noether's first theorem which leads to the divergence relationships in question. However, given that the same continuity equations and conservation laws can often be derived by several means, the same should also apply to other means leading to the same results. In particular, this should then apply to the derivations proceeding via Noether's second theorem where these lead to the same results as those obtainable via Noether's first theorem. An example is the case of classical electromagnetism with matter, where as Brading and Brown [2000] show the conservation of electric current and charge follows from Noether's first and second theorems alike. Therefore, we should be able to find $C$ also in the general expression for conservation laws obtainable via the extended Noether's second theorem. Unfortunately, it is obscure from Noether's and Bessel-Hagen's formulations of the dependences arising from that theorem where $C$ is to be found there, as it does not figure in Bessel-Hagen's formulation explicitly: indeed his equation (10) expressing the dependences it is identical to Noether's corresponding equation (16), as mentioned[4].

Besides this another way in which $C$ affects Noether's theorems and conservation laws can be found in Bessel-Hagen's comment on the converse of Noether's first theorem, i.e. on the derivation of finite symmetries from the divergence relationships. Namely, Bessel-Hagen says that in the extended case there is an ambiguity about how to decompose $B$ into $C$ and the rest. To remind, the rest here per Bessel-Hagen's' equation (6) consists of $A$ and $\delta f$, while $f$ is here the integrand, in our case a Lagrangian or a Lagrangian density. In other words, it follows from what Bessel-Hagen says that if one gets the divergence relationships with certain $C$, $A$ and $\delta f$, one may try get back to finite symmetries using another such quantities by redistributing parts of the previous $C$ into $A$ and $\delta f$, or vice versa. Bessel-Hagen formulates some constraints under the conjunction of which one is guaranteed to get finite symmetries when going back, but there are possibly other constraints which allow the same. Specifying these would allow to evaluate the amount of freedom in the choice of $C$, $A$ and $\delta f$, and thus indirectly of $f$ itself. But the general point is anyway clear. Namely, the same dependence relationships, and hence the same continuity equations and conservation laws, may thus correspond to many finite symmetries and to many $f$'s, i.e. Lagrangians or Lagrangian densities. Examples of different Lagrangian densities leading via Noether's first theorem to the same conservation laws, though presumably pathological ones, can be found in [Brown and Holland 2004a, Sect. III].

---

4    Except for a typo: Bessel-Hagen's $\nu$ should be $\sigma$.

# Physical aspects

A warning should be issued: in this part my discussion is somewhat more technical than in the previous two.

## *Preliminaries on the n-body problem and classical electromagnetism*

Passing to the application of extended Noether's theorems to physics, Bessel-Hagen discusses two cases: the *n*-body problem (§2) and classical electromagnetism (§§3-7). To begin with, some preliminary details on these cases are worth mentioning.

The *n*-body problem goes back to Isaac Newton (years of life 1643-1727 by the Gregorian calendar) and consists in describing the evolution of *n* massive bodies, which presupposes being able to account for their mutual gravitational forces. Though reputed as difficult to be solved in practice, this problem has a well-defined general description in terms of Hamilton's mechanics and Newton's gravity. It is apparently this description that Bessel-Hagen uses in his §2, as we can see from his recurrence to momenta $\dot{x}_i$ and to the gravitational constant $\kappa$ (in contemporary notation, G) respectively. Here we have a particle theory, and Bessel-Hagen refers to the conservation laws obtained for it as "first integrals".

Classical electromagnetism, meanwhile, is a field theory describing electromagnetic fields and their interactions, formulated by James Clerk Maxwell (1831-1879) who combined electricity and magnetism in his equations. This theory has a free version, with electromagnetic fields in vacuum (or "aether", as Bessel-Hagen says), and an interaction version, with electromagnetic fields in the presence of massive matter. Bessel-Hagen briefly states Maxwell's equations for the vacuum version as his equations (14) from §4, denoting the components of the electromagnetic field by $f_{ik}$ (the contemporary notation is $F_{\mu\nu}$). The same theory whether with or without matter also has a formulation in terms of the electromagnetic potential, whose components Bessel-Hagen denotes by $\varphi_i$ (the contemporary notation is $A_\mu$). The relationship between the two formulations is given by Bessel-Hagen's equation (15) from §4 (in the contemporary notation it says $F_{\mu\nu} = \partial_\mu A_\nu - \partial_\nu A_\mu$).

The desire to make Maxwell's equations manifestly covariant under Lorentz transformations led Einstein to the formulation of special relativity in 1905. This resulted in turn, with Minkowski's help, in a unified treatment of space and time. This change got also reflected in the notation: instead of denoting spatial and temporal variables completely differently as before, both alike got denoted by *x* followed by an index from the same continuous range. (The contemporary indexing ranges from 0 to 3, with 0 designating time and 1-3 space dimensions. Arbitrary dimensions are denoted by Greek letters like $\mu$ and $\nu$ if they concern the 0-3 range, by Latin letters like $i$ if they concern the 1-3 range.)

When discussing classical electromagnetism in §§4-7, Bessel-Hagen uses both kinds of notation. In §3 he gives a table with partial correspondence between them for some electromagnetic terms. He calls these a 4-dimensional notation (this is roughly a relativistic notation but ranging from 1 to 4) and a 3-dimensional notation, which I will name 3+1-dimensional (this is roughly a pre-relativistic notation with space dimensions denoted by $x$, $y$, $z$, while time is denoted by $t$ and is taken to occur in the combination $ict$; in that notation spatial and temporal components of 4-dimensional quantities get denoted by distinct variables).

One reason for him using the 4-dimensional notation apparently has to do with the distinction between real and imaginary quantities, the imaginary $i$ figuring in the time coefficient of the 3+1 dimensional notation but not in the 4-dimensional one. Thus in the introduction Bessel-Hagen notes that one has to pass to the latter notation when considering the conformal group, that group being defined on Minkowski space $\mathbb{R}^4$ where each dimension is indexed by the reals ($\mathbb{R}$). In §4 he further links the question whether the parameters of that group are real with the physical point of view on conservation laws as opposed to their formal discussion. In §7, on the other hand, he says that it is the 3+1-dimensional notation that allows to exhibit the physical meaning of the electromagnetic conservation laws. The validity of this last reason may be questioned, however, as currently one has no problem of making sense of conservation laws despite the relativistic 4-dimensional notation being common.

### *Bessel-Hagen's conservation laws for the **n**-body problem*

In §2 Bessel-Hagen derives 10 "first integrals" for the $n$-body problem. As he remarks, these are already known and their derivation using "Lie's methods" was already made in [Engel 1916]. His own novelty however is to exhibit a derivation using Noether's theorems and therefore to illustrate their usefulness in a simple context.

As the symmetry group he considers is finite, we note that he is actually going to use Noether's first theorem alone. Unsurprisingly, the translation [1921/2006] mentions a single theorem, but in the original the plural form is used. Bessel-Hagen promises to give the calculations completely so as to provide an analogy with the electromagnetic case. In fact his presentation is neither complete nor exactly analogous, but it is quite detailed and quite similar nevertheless.

In contemporary terms what Bessel-Hagen does in §2 is the following.

First of all he specifies the action from which the relevant Euler-Lagrange equations for the *n*-body problem follow via Hamilton's principle, as well as a Lagrangian figuring in that action together with the dependent variables ($x_{ik}$) and the independent variable (*t*) this Lagrangian is a functional of. Next Bessel-Hagen considers the finite 10-parameter inhomogeneous Galilean group consisting of time translations, space translations, spatial rotations and Galilean boosts, and lists the variations in the independent and the dependent variables brought about by these transformations under respectively (11 a), (11 b), (11 c) and (11 d)[5].

He notes moreover that while the relevant equations are invariant under all the transformations concerned, the Lagrangian (he should have said the action, as this is what he defined symmetries up to a divergence for) is invariant under the first three transformations but invariant up to a divergence under the last one. He therefore gives explicitly the variation in the Lagrangian for that last (Galilean boosts) case and denotes that variation by Div*C*. Afterwards he gives the explicit form of the equations (5) and (8) from §1 for his particular case. To remind, these specify respectively the total variation in the dependent variables and the to-be-conserved quantity *B*. In the expression for *B*, as expected, the term *C* again figures.

Next in (12 a-d) the divergence relationships arising from Noether's first theorem are given in the same order as the variations (11 a-d) caused by the corresponding transformations. Then components of the Euler-Lagrange expressions $\psi$ are specified and are next set to 0. Under this supposition the left-hand sides of (12 a-d) vanish, leading according to Bessel-Hagen to the following conservation laws, in the same order as before: a) "the energy theorem", b) "the first three centre-of-mass theorems (also known as the momentum theorems)", c) "the three area theorems", d) "the three second centre-of-mass theorems". In contemporary terms these correspond to the conservation of energy, linear momentum, angular momentum and centre of momentum respectively.

Finally, on the case d) Bessel-Hagen says that it differs from the usual formulation but that it can be obtained by noting that by (12 b) the right-hand side of the latter equation equals to a constant $c_k$, whence follows the equation (13). This one says that the first term in the brackets of the right-hand side of (12 d) equals to $c_k t + c_k'$, where $c_k'$ is another constant. He says however that he prefers (12 d) because this inserts the conservation laws concerned into the sequence of others and because this is going to provide an analogy to the electromagnetic conservation law (28).

---

### *Comments on the Galilean boost conservation laws for the* **n**-*body problem*

Personally I have two questions to Bessel-Hagen's treatment of the *n*-body problem, both related to the (12 d) case corresponding to Galilean boosts: What does the explanation at the end of §2 means, the one where he mentions the usual formulation and the constants? And also, as (12 d) is the divergence relationship for Galilean boosts under which the Lagrangian acquires a change Div$C$, where is $C$ in (12 d) itself?

Addressing the latter question first, what follows from our discussion of §1 is that when the Lagrangian acquires a change Div$C$ under a transformation, $C$ should appear as part of $B$ on the right-hand side of the divergence relationship corresponding to that transformation. Now in the divergence relationship (12 b) corresponding to Galilean boosts there is a sum of two terms on the right-hand side, therefore these terms compose $B$ and perhaps one of these terms is $C$. On the other hand, the general definition for $B$ in the *n*-body problem case is given just above (12), and there 3 terms feature: $C$, a term involving a variation in $x_{ik}$, and a term involving a variation in $t$. Substituting in the last two terms the variations from (11 d), we see that the last term vanishes while the other term becomes the second term on the right-hand side of (12 d). Therefore the first term on the right-hand side of (12 d) should be the sought $C$.

And indeed, this is almost exactly confirmed by the definition of $C$ just below (11). The only major discrepancy is that in that definition of $C$ as well as in (11 d) we see some $\gamma_k$ which is absent from (12 d). The explanation is that this $\gamma_k$ is apparently a group parameter for Galilean boosts, in the same way as $\beta_{k\rho}$, $\alpha_k$ and $\tau$ from respectively (11 c), (11 b) and (11 a) are apparently group parameters for the other transformations concerned. Such group parameters should figure indeed in the variations like those expressed in (11) but not in the divergence relationships like those expressed in (12): how this happens is explained in [Brading and Brown 2000, just below their equation (9)] and has to do with the fact that these are parameters and hence constants. Another discrepancy is the absence of summations over $k$ and $\rho$ in (12) as compared to (11). Of this the separation with respect to the parameters, expressed in the equations just above Bessel-Hagen's equation (7), seems to be the reason.

Now what about the other question involving Bessel-Hagen's claim that (12 d) is unusual in some respect? Arguably, the presence of $C$ there is already an unusual feature. However, it actually seems usual for boosts (see e.g. [Brading and Brown 2000, n. 4] and [Brown and Holland 2004a, the end of Appendix A]). Thinking therefore of another explanation, it seems to me that the usual conservation laws Bessel-Hagen is seeking should have the form not of Div$B = 0$ but of $B$ = const. One argument for this is that Engel gives conservation laws in the latter form at the end of his [1916] on which Bessel-Hagen's §2 relies a lot, including in its last part where the unusual character of the conservation law corresponding to what Bessel-Hagen denotes by (12 d) is highlighted. If my interpretation is right, then Bessel-Hagen's unnumbered equation towards the end of §2 which expresses the $B$ from the right-hand side of (12 b) as a constant $c_k$ exemplifies the usual conservation law Bessel-Hagen is seeking, while the equation (13) is his attempt at putting in the same form the term $C$ from (12 d). As (13) should therefore express for him a usual conservation law corresponding to (12 d), I will denote it as (13 d) and will introduce the designations (13 c), (13 b) and (13 a) for Bessel-Hagen's usual conservation laws corresponding respectively to (12 c), (12 b) and (12 a). This is for convenience when discussing the electromagnetic case later on, as Bessel-Hagen promises an analogy with the $n$-body problem case there.

### Bessel-Hagen's conservation laws for classical electromagnetism, with comments

In contemporary terms Bessel-Hagen's discussion of classical electromagnetism (§§3-7) proceeds as follows: after the overview of the notations in §3 he presents the symmetries of free classical electromagnetism in §4, a dependence and divergence relationships for the free case in §5, divergence relationships for classical electromagnetism with matter in §6, and continuity equations as well as conservation laws in part for the matter case and in part for the free case in §7. Let us discuss §§4-7 in some more detail.

As the title of §4 says, this section is supposed to be on symmetries of Maxwell's equations, but while reading it we see that it is symmetries of an electromagnetic action that are presented, just as is needed for applying Noether's theorems. The variation considered is however narrower than is needed for the theorems, and reminds of Hamilton's principle instead in that the transformations are supposed to act on the bulk of the domain of integration but not at its boundaries. So apparently here Bessel-Hagen's desire to spell out the link of relevant symmetries with Maxwell's equations gets in contradiction with his desire to prepare the ground for the application of Noether's theorems. In fact, however, the two need not conflict: it would suffice to say that the same action allows to derive Maxwell's equations via Hamilton's principle for a restricted class of variations and to derive conservation laws via Noether's theorems for a larger class of variations.

The way Bessel-Hagen introduces Maxwell's equations seems to only add to the confusion: in (14) these are presented for the electromagnetic field and its dual, but then the constraint (15) is introduced on the link between the electromagnetic field and the electromagnetic potential (in the contemporary notation, $F_{\mu\nu} = \partial_\mu A_\nu - \partial_\nu A_\mu$) with the result that the dual equations then get eliminated from the discussion and the other equations get modified. Arguably a less confusing way would be to start with the modified equations directly, and to introduce (15) at the end of the section where Bessel-Hagen currently mentions it again.

Perhaps the most important information from §4 is the list of symmetries of the electromagnetic action Bessel-Hagen considers. These consist of the finite 15-parameter conformal group and of the infinite group involving a single function. The latter symmetry is apparently the usual gauge symmetry, in contemporary terms $A_\mu(x) \rightarrow A_\mu(x) + \partial p(x)/\partial x_\mu$ for a function $p(x)$. It is when discussing this symmetry that Bessel-Hagen mentions (15) once again. Of the former group Bessel-Hagen was saying already in the introduction that it consists of the (inhomogeneous) Lorentz group together with 5 more transformations; I will name these a scale transformation and 4 special conformal transformations following [Kastrup 2008]. In §4 Bessel-Hagen adds to this in particular that the conformal group induces variations of the electromagnetic potential. This is as expected, for the transformations of the conformal group act on the spatiotemporal variables, while the electromagnetic potential variable is dependent on them. Therefore, one will have to take into account this induced variation when deriving the divergence relationships for conformal transformations.

Such derivations occupy most of §5, in which Bessel-Hagen aims at deriving the 15 divergence relationships corresponding to the application of Noether's first theorem to the conformal group and the single dependence corresponding to the application of Noether's second theorem to the gauge symmetry.

To do so he firstly lists the components of the relevant Euler-Lagrange expressions (16), the proper variations of the independent variables (17 a-d) and the induced variations of the dependent variables (18) arising from the conformal group, as well as the proper variations of the dependent variables arising from the gauge transformation (just below 18). He then says that in the gauge transformation case the total variation of the dependent variables (5) reduces to this last variation, and that therefore (9) and (10) yield (19). As (9) and (10) are the equations Bessel-Hagen introduced in §1 when discussing Noether's second theorem, we recognise in (19), though unfortunately without an explicit indication from his side, the sought dependence arising from the gauge symmetry. This dependence says that the divergence of the electromagnetic Euler-Lagrange expressions vanishes identically. Bessel-Hagen does not comment on $C$ here either, but we do not expect it in (19), as the gauge symmetry should be exact for the relevant Lagrangian density. Therefore (19) could be derived just as well using Noether's second theorem in its non-extended form.

Passing to Noether's first theorem, Bessel-Hagen goes as far as obtaining an expression for $B$ (20) using (8). The predicament is however that upon substitution of the variations (17 a-d) and (18) there the divergence relationships he gets come out as long, complicated and featuring the electromagnetic potential instead of the electromagnetic field alone. To avoid this Bessel-Hagen adds the gauge symmetry under consideration and choses its function $p(x)$ to be a multiple of the variation (17) with the electromagnetic potential. This choice as he notes allows to eliminate a similar expression from $B$ as specified by (20), but in fact, as we may note, it also eliminates the variation (18) altogether. That is, his choice of the proper variation of the electromagnetic potential serves to annihilate the induced variation of the electromagnetic potential. The result are nicely looking divergence relationships (22) featuring the variation in the independent variables (17) alone. (23 a-d) are the particular cases of (22) for the different transformations constituting the conformal group.

What is striking about this derivation of Bessel-Hagen's is what he does with the variations in the electromagnetic potential there. Either he is wrong about introducing the gauge symmetry and the specific choice of its function, but then the question what to do with the unnatural expression for $B$ (20) returns; or he is right in doing what he does, but then the question is why one would be allowed to combine the proper and the induced variations of the electromagnetic potential in the way he does.

Passing to the short §6, Bessel-Hagen says there that in order to bring in the physical significance he is going to make the Euler-Lagrange expressions vanish in the vacuum case and make them equal to some $P$ in the matter case. So far, to remind, he was discussing the vacuum case alone, so in fact this amounts to extending the discussion to the matter case as well. Equating the Euler-Lagrange expressions with $P$ for that case may seem mysterious: what is $P$, were did Bessel-Hagen took the resulting equation at all, and how are we going to get conservation laws if the Euler-Lagrange expressions are not going to vanish? The translation here is very misleading, designating $P$ as "flow" instead of "current". Understanding $P$ in the latter way, we see however that this is the same as $J$, the familiar electric current to be conserved. Meanwhile, equating the gauge Euler-Lagrange expressions with $J$ is an instance of the 'Coupled Field Equations' from [Brading and Brown 2000], which hold for interaction theories with infinite symmetry groups. And, finally, as [ibid.] shows the Euler-Lagrange equations will have to vanish in the matter case too if we are to get conservation laws there. But Bessel-Hagen does not go as far, switching back to the free case in §7. As a result, he picks the dependence (19) which is valid for the free case and is saying that the divergence of the gauge Euler-Lagrange equations vanishes identically. This and the 'Coupled Field Equations' allow him to state in (26) that $J$ is also conserved identically in the free case. As Brading shows [2002, Appendix], though without mentioning Bessel-Hagen, this result actually relies on the extended Noether's second theorem rather than on the original one. It is a pity that Bessel-Hagen does not highlight this himself.

Instead, Bessel-Hagen concentrates on the conservation laws arising from the conformal symmetries. In §6 he tries to obtain them for the matter case out of the divergence relationships (23 a-d). This succeeds for (24 a) and (24 b), where the right-hand sides vanish, but not for the scale (24 c) and special conformal (24 d) transformations, where these sides happen to feature the matter energy-momentum tensor instead. The same holds after (24) gets changed into (25) due to Bessel-Hagen's switching in §7 to the 3+1-dimensional notation (where $r$ concerns space and $z$ time). So he continues by reverting back to the vacuum case, and integrates over the currents. As a result, he finally gets conservation laws (27 $a_r$, $a_z$, $b_r$, $b_z$, c, $d_r$, $d_z$) for all of the conformal transformations.

All of these are expressed in the form Q = const., confirming my hypothesis mentioned when discussing §2 by which this kind of form corresponds to the usual conservation laws in Bessel-Hagen's (and Engel's) sense. More precisely, (27 $a_r$, $a_z$, $b_r$) correspond respectively to what I called (13 b, a, c), while Bessel-Hagen's discussion of (27 $b_z$) is almost completely analogous to his discussion of what I called (13 d) (ironically, he adds relativistic considerations there despite using the 3+1-dimensional notation). Given that analogy, we should be able to find $C$ in (27 $b_z$) as well. Again, Bessel-Hagen is completely silent on that, but the reasoning analogous to the one I presented when analysing (13 d) shows that the left-hand side of (27 $b_z$) is precisely the $C$ sought. Therefore (27 $b_z$), as much as (13 d), are for Bessel-Hagen conservation laws for $C$ alone. This is in contrast with the more recent approach of e.g. Brown and Holland, who understand $C$ as part of a conserved current and of a conserved charge [2004a, p. 35, equations (8)-(11)]. They discuss conservation laws arising from variations of the dependent variables alone, but Noel Doughty [1990, pp. 338-339] places $C$ (his $\mathbf{\Delta} A_\mu$) into the current and charge when discussing the variations of the dependent and independent variables alike.

More mysterious is the situation with Bessel-Hagen's remaining conservation laws (27 c, $d_r$, $d_z$), which correspond to the scale and special conformal symmetries and do not have analogues in the $n$-body problem case. By that reason it is not clear in particular whether $C$ is present in them, but also it is not clear more generally how to interpret these remaining conservation laws and the quantities conserved in them. Bessel-Hagen's own proposal at the end of his article is to introduce new notions for the charges figuring in these conservation laws. Klein as to him was also concerned with the interpretation of these results, so soon after Bessel-Hagen submitted the article to him in view of the publication Klein wrote to Pauli asking him among other things to assist with the interpretational question (Klein's letter is translated in [Kosmann-Schwarzbach 2011, pp. 159-160]). And at the same time, as Klein mentions [ibid.], Bessel-Hagen was going for vacations to Berlin and was supposed to ask the same question to Max Plank who he knew in person, apparently from his studies of physics preceding the Göttingen move. In the published version, however, Bessel-Hagen's acknowledgements at the end of his article go to Noether and to the physicist Paul Hertz alone, which suggests that neither Plank nor Pauli were able to clarify the interpretational question if of course they considered it at all.

More recently Plybon [1972] returned to the question of interpreting Bessel-Hagen's mysterious conservation laws. According to his analysis (with equation numbers replaced by Bessel-Hagen's), (27 a) follow from Maxwell's equations and the definition of the energy-momentum tensor [ibid., equation (13)], while (27 b, c) follow from (27 a) plus some properties of the same tensor, and none of (27 a, b, c) "can be derived from the remaining two without further information" [ibid., p. 1001], while (27 d) can be derived from the other three without further information; hence for Plybon (27 d) are unphysical, while for the interpretation of (27 c) he advises to consult [Jackiw 1972], who discusses then recent uses of scale symmetry in physics.

One problem with Plybon's account is however that he considers the free electromagnetism case alone, while one may wonder whether any conservation laws have any sense in the absence of matter. A further problem is that taking special conformal transformations as unphysical and scale symmetries as physical is implausible, for there exists the link between the two, and the link between the conservation laws arising from the two, as discussed and advocated by Kastrup [2008, pp. 663-666]. In fact this link is most clear precisely from the presence of the matter energy-momentum tensor on the right-hand sides of both (24 c) and (24 d) (cf.[ibid.] and [ibid., pp. 656-657]). Plybon's insistence on the free case obscures this link, as does Bessel-Hagen's. But if one returns to the matter case, then again because of the terms on the right-hand sides of (24 c) and 24 d) conservation laws for the corresponding transformations in general do not obtain, as Bessel-Hagen rightly notes (see likewise [ibid.]). In other words, if Kastrup is right, then all of Bessel-Hagen's mysterious conservation laws would have to have a physical sense, and not only one of them as Plybon was proposing. However, perhaps no sense is to be sought for any of them at all, at least not within classical electromagnetism, given that in the case where that sense could have arisen, namely in the matter case, these conservation laws in general do not hold.

**Conclusion**

Bessel-Hagen wrote his article in his young years, contributing to Klein's late Erlangen programme and building particularly on Noether's results. His work connected Noether's theorems more straightforwardly with physics and extended their domain of application. The term $C$ which arises from the extension of Noether's theorems to symmetries up to a divergence contributes in the quantities conserved. Such quantities can be found in conservation laws for boosts in the $n$-body problem and the electromagnetic case: Bessel-Hagen's discussion contains enough information to infer that, though he does not carry out the relevant analysis explicitly. Meanwhile, his conservation laws for scale and special conformal symmetries perhaps do not have any physical significance within the electromagnetic theory, because in that context they are only guaranteed to hold in the absence of matter. Nevertheless, this does not preclude them from having such a significance within other physical theories.

**References**

*Historical references*

theory of the spatially closed world (trans.: Chen, Chiang-Mei, Nester, James M., and Vogel, Walter). Available at https://arxiv.org/abs/2006.14743

*Scholarly references*